\documentclass[preprint,eqsecnum,showpacs,floatfix]{revtex4}
\usepackage{graphicx}
\usepackage{epsf}
\usepackage{bm}
\usepackage[hypertex]{hyperref}
\def\he#1{$^{#1}$He}

\def\bra#1{\bigl\langle {#1}\bigr|}
\def\ket#1{\bigl| {#1} \bigr\rangle}
\def\rvec {{\bf r}}

\def\qvec {{\bf q}}

\newcommand{\I}    [0] {{\rm i}}            

\newcommand{\Vph} [1]
           {{\widetilde V_{\!\scriptscriptstyle\rm p_{\bar{\ }\!}h\!}(#1)}}
\newcommand{\SF}      [1] {S_{\rm\scriptscriptstyle F}(#1)}

\begin{document}

\title{Theoretical analysis of neutron and X-ray scattering data on $^3$He}
\author{E.\ Krotscheck$^{\dagger +}$ and M. Panholzer$^\dagger$}
\affiliation{$^\dagger$Institut f\"ur Theoretische Physik, Johannes
Kepler Universit\"at, A 4040 Linz, Austria}
\affiliation{$^+$Department of Physics, University at Buffalo SUNY
Buffalo NY 14260}

\begin{abstract}
  X-ray scattering experiments on bulk liquid \he3
  \cite{Albergamo,Albergamo_comment,Albergamo_reply} have indicated
  the possibility of the existence of a sharp collective mode at large
  momentum transfers. We address this issue within a manifestly
  microscopic theory of excitations in a Fermi fluid that can be
  understood as proper generalization of the time-honored theory of
  Jackson, Feenberg, and Campbell
  \cite{JacksonSkw,FeenbergBook,Chuckphonon} of excitations in
  $^4$He. We show that both neutron and X-ray data can be well
  explained within a theory where the high momentum excitations lie in
  fact inside the particle-hole continuum.  ``Pair fluctuations''
  contribute a sharpening of the mode compared to the random phase
  approximation (RPA). When the theoretical results are convoluted
  with the experimental resolution, the agreement between theory and
  X-ray data is quite good.
\end{abstract}
\maketitle
\section{Introduction}

The helium fluids \he3 and \he4 are the prime examples of strongly
correlated quantum many-body systems. Governed by a simple
Hamiltonian, yet very dense, they have been studied for decades and
still offer surprises leading to new insights. It is fair to say that
understanding the helium fluids lies at the core of understanding
other strongly correlated systems.

Liquid \he3 is for many reasons the more challenging substance for
both, theoretical and experimental investigations.  Key information on
the dynamics of the system is provided by the dynamic structure
function $S(q;\omega)$.  Experimentally, the dynamic structure
function of \he3 is mostly determined by neutron scattering. The
results are well documented in a book \cite{GlydeBook}, the
theoretical and experimental understanding a decade ago has been
summarized in Ref.~\onlinecite{GFvDG00}.  Recent inelastic X-ray
scattering experiments have led to a controversy on the evolution of
the zero sound mode at intermediate wave-vectors
\cite{Albergamo,Albergamo_comment,Albergamo_reply}. It is one of the
purposes of this paper to examine this issue on the basis of
microscopic many body theory.

The ground state of liquid \he4 is well understood from both
diagrammatic many-body techniques and simulation methods, see
Ref. \onlinecite{KroTriesteBook} for a collection of pedagogical
articles. Building upon pioneering work by Jackson, Feenberg, and
Campbell \cite{JacksonSkw,FeenbergBook,Chuckphonon}, recent
developments \cite{eomI,QFS09_He4,eomII} have brought manifestly
microscopic theories of \he4 to a level where quantitative predictions
for the excitation spectrum are possible up to momentum transfers well
beyond the roton minimum.

We have recently generalized the theory of Refs.
\onlinecite{JacksonSkw,FeenbergBook,Chuckphonon} to Fermions
\cite{2p2h}.  From the theoretical point of view, the major technical
problem is to deal with the multitude of exchange effects in an
efficient manner; we have formulated a tractable strategy for
including pair-excitation effects in Fermi fluids. This paper is
concerned with a theoretical study of excitations and damping in
normal liquid \he3 at atomic wave lengths based on that theory.

\section{Dynamic Many-Body Theory}

\subsection{Equations of Motion}

This section gives a very brief compilation of the techniques
developed in Ref. \onlinecite{2p2h}. We deliberately refrain from any
explanation and refer to the original work and earlier pedagogical
material \cite{KroTriesteBook} for details.

Microscopic many-body theory starts from a phenomenological
Hamiltonian for $N$ interacting fermions,
\begin{equation}
H = -\sum_i\frac{\hbar^2}{2m}\nabla_i^2 + \sum_{i<j}
v\left(\left|\rvec_i\!-\!\rvec_j\right|\right) \;.
\label{eq:Hamiltonian}
\end{equation}
For strong interactions, correlated basis functions (CBF)
theory \cite{FeenbergBook} has proved to
be an efficient and accurate method for obtaining ground state
properties. It starts with a variational wave function of the form
\begin{equation}
  \vert \Psi_{\bf o} \rangle = {F \; \vert \Phi_{\bf o} \rangle \over
  \langle \Phi_{\bf o} \vert\,  F^{\dagger} F^{\phantom{\dagger}}
    \vert \Phi_{\bf o} \rangle^{1/2} } \;,
\label{eq:JastrowWaveFunction}
\end{equation}
where $\Phi_{\bf o}(1,\ldots,i,\ldots,N)$ is a model state, normally a
Slater--determinant, and ``$i$'' is short for both spatial and $\nu$
discrete (spin and/or isospin) degrees of freedom. The {\it
  correlation operator}\/ $F(1,\ldots,N)$ is suitably chosen to
describe the important features of the interacting system. Most
practical and successful is the Jastrow--Feenberg \cite{FeenbergBook}
form
\begin{eqnarray}
	F(1,\ldots,N) &=&
	\exp\left\{{1\over2}\left[\sum_{1\le i<j\le N} u_2({\bf r}_i,{\bf r}_j)
	+ \sum_{1\le i<j<k\le N}u_3({\bf r}_i,{\bf r}_j,{\bf r}_k)
	+ \ldots\right]\right\} \,.
\label{eq:JastrowCorrelations}
\end{eqnarray}
The $u_n({\bf r}_1,\ldots,{\bf r}_n)$ are made unique by requiring
them to vanish for $|{\bf r}_i\!-\!{\bf r}_j| \to\!\infty$
(``cluster property''). 

From the wave function (\ref{eq:JastrowWaveFunction}),
(\ref{eq:JastrowCorrelations}), the energy expectation value
\begin{equation}
\label{eq:energy}
H_{\bf o,o}
\equiv \left\langle \Psi_{\bf o} \right| H \left| \Psi_{\bf o}\right \rangle
\end{equation}
can be calculated either by simulation or by integral equation
methods. The hierarchy of Fermi-Hypernetted-Chain (FHNC)
approximations is compatible with the optimization problem, {\it
  i.e.\/} with determining the optimal {\it correlation functions\/}
$u_n({\bf r}_1,\ldots,{\bf r}_n)$ through functionally minimizing the
energy
\begin{equation}
{\delta H_{\bf o,o}\over \delta u_n({\bf r}_1,\ldots,{\bf r}_n)} = 0 \,.
\label{eq:Euler}
\end{equation}

The dynamics of the system is treated with the logical generalization
of the wave function (\ref{eq:JastrowWaveFunction}),
(\ref{eq:JastrowCorrelations}) by writing the response of the system
to a weak and time dependent external field $H_{\rm ext}(t) = \sum_i
\delta h_{\rm ext\/} ({\bf r}_i;t)$ in the form
\begin{eqnarray}
\left|\Psi(t)\right\rangle\; &=&\displaystyle\frac{1}{\sqrt{\cal N}}\,
\displaystyle e^{-\I E_0 t/\hbar}
        \left|\Psi_0(t)\right\rangle\,,\nonumber\\
        \left|\Psi_0(t)\right\rangle &=&
        F\, e^{{1\over 2}\delta U(t)}
                \left|\Phi_0\right\rangle
        \quad,\quad {\cal N}\equiv
        \left\langle\Psi_0(t) \!\mid\! \Psi_0(t)\right\rangle \,.
\label{eq:PsiExcOft}
\end{eqnarray}
where $F$ is taken from the ground state calculation, and $\delta
U(t)$ is a sum of $n$--particle--$n$--hole excitation operators:
\begin{equation}
\delta U(t) = \sum_{ph}\delta u^{(1)}_{ph}(t)\,
              a^\dagger_p a^{\phantom{\dagger}}_h +
              \textstyle\frac{1}{2}\displaystyle\!\!
              \sum_{pp'hh'}\! \delta u^{(2)}_{pp'hh'}(t)\,
              a^\dagger_p a^\dagger_{p'}
              a^{\phantom{\dagger}}_{h'}a^{\phantom{\dagger}}_h
+ \ldots\ ;
\label{eq:UFermiOft}
\end{equation}
As a convention, we will label ``particle'' (unoccupied) and ``hole''
(occupied) states with $p,p'$ and $h,h'$, respectively. Discrete
degrees of freedom are suppressed.

Equations of motion for the $\delta u^{(i)}(t)$ are derived from
the linearized action principle \cite{KramerSaraceno,KermanKoonin}
\begin{equation}
\delta \int\! dt\, {\cal L}(t)
        = \delta \int\!\! dt \left\langle\Psi(t)\left|H +
        H_{\rm ext\/} (t)
        -\I\hbar{\partial\over\partial t}
        \right|\Psi(t)\right\rangle = 0\,.
\label{eq:Action}
\end{equation}
The analytic and diagrammatic manipulations to bring the resulting
equations of motion into a numerically tractable form are lengthy and
delicate \cite{2p2h}, but the result is surprisingly simple and can be
cast into the familiar form of the time-dependent Hartree-Fock (TDHF)
theory \cite{ThoulessBook} with  {\em energy dependent\/} effective
interactions.  The induced density fluctuations is expanded in terms
of matrix elements of the density operator in the non-interacting
system, $\rho^F_{0,ph}({\bf r}) = \bra{h}\delta\hat\rho({\bf r})\ket
{p}$.
\begin{eqnarray}
\delta\rho({\bf r};\omega) 
	&=& \frac{1}{2}\sum_{ph} \left[
          \rho_{0,ph}^{\rm F }({\bf r})\,
          \delta c^{(+)}_{ph}(\omega) +
          \rho_{0,ph}^{\rm F*}({\bf r})\,
          \delta c^{(-)}_{ph}(\omega)\right]
\;.
\label{eq:d_rho_r_om}
\end{eqnarray}
The amplitudes $\delta c^{(\pm)}_{ph}(\omega)$ are related by the THDF
equations to the matrix elements $h_{ij} = \bra{i}\delta h_{\rm
  ext\/}\ket{j}$ of the external field.

\begin{equation}
\left(
\begin{array}{cc}
(\hbar\omega -e_{ph}) \delta_{p,p'}\delta_{h,h'} +
V^{(A)}_{ph,p'h'}(\omega)& V^{(B)}_{pp'hh',0}(\omega) \\
 V^{(B)}_{0,pp'hh'}(\omega) &
-(\hbar\omega +e_{ph})\delta_{p,p'}\delta_{h,h'}+V^{(A)}_{p'h',ph}(\omega)
\end{array}\right)
\left(
\begin{array}{c}
\delta c_{ph}^{(+)}\\ \delta c_{hp}^{(-)}
\end{array}\right)
=
2\left(
\begin{array}{c}
h_{ph}\\h_{hp}
\end{array}\right)\,,
\end{equation}
where $e_{ph}=t(p)-t(h)$ is the particle-hole excitation energy of
the non-interacting system and $t(q) = \hbar^2 q^2/2m$ is the kinetic
energy.

Thus, the machinery of microscopic many-body theory has led to a
definition of the effective interactions $V^{(A)}_{ph,p'h'}(\omega)$
and $V^{(B)}_{pp',hh'}(\omega)$ in terms of the correlation operator
$F$ and the underlying microscopic Hamiltonian.

\subsection{Local interactions}

Much of our qualitative understanding of the dynamics of interacting
quantum fluids is based on the ``random phase approximation'' (RPA).
In that approximation, the density-density response function is
of the form
 \begin{equation}
   \chi(q; \omega) = {\chi_0(q; \omega)\over 1 - \Vph{q}\,
   \chi_0(q; \omega)}
  \label{eq:RPAresponse}
  \end{equation}
Here, $\chi_0(q; \omega)$ is the
Lindhard function, and $\Vph{q}$ an appropriately defined static
``particle--hole interaction'' or ``pseudo-potential \cite{Aldrich}''.
It is related to the dynamic structure function
\begin{equation}
S(q;\omega)=-\frac{\hbar}{\pi}\Im m[\chi(q;\omega)]\,\theta(\omega)
\label{eq:FD}
\end{equation}
which satisfies, amongst others, the
sum rules
\begin{eqnarray}
 m_0(q)=S(q)&=&\int_0^{\infty}d\hbar\omega \;S(q; \omega),
 \label{eq:m0}\\
 m_1(q)=\frac{\hbar^2q^2}{2m}&=&\int_0^{\infty}d\hbar\omega \;\hbar\omega \,
 S(q; \omega)
 \label{eq:m1} \,,
\end{eqnarray}
where $S(q)$ is the static structure factor. A version of the RPA with
an effective interaction can be derived by restricting the excitation
operator (\ref{eq:UFermiOft}) to one-particle-one hole amplitudes, and
omitting (or approximating) exchange terms.

\begin{figure}
\includegraphics[width=0.48\textwidth,angle=270]{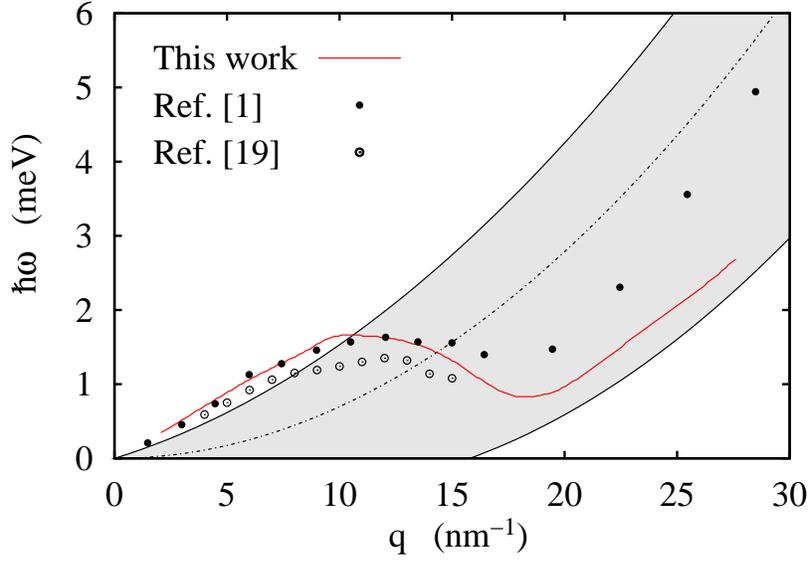}
\caption{(color online) A schematic picture of the dynamic structure
  function $S(q;\omega)$ in \he3. The gray shaded area is the
  particle-hole continuum of non-interacting particles, the circles
  show the maximum intensity of neutron-scattering data
  (Ref. \onlinecite{Fak94}) and the dots are X-ray scattering data
  \cite{Albergamo}. The red line shows the results of our calculation.
  See also Sec. \ref{sec:Skw} for further discussions.}
\label{fig:ezsXray}
\end{figure}

The RPA (\ref{eq:RPAresponse}) for $\chi(q; \omega)$ displays the
essential features of the dynamic structure function $S(q;\omega)$
qualitatively correctly: $S(q;\omega)$ can be characterized as being a
superposition of a collective mode similar to the phonon-maxon-roton
in $^4$He, {\it plus\/} an incoherent particle-hole band which
strongly dampens this mode \cite{PinesPhysToday} where this is
kinematically allowed. A schematic drawing is shown in
Fig. \ref{fig:ezsXray}.

The picture misses important physics: In \he3 the RPA, when defined
through the form (\ref{eq:RPAresponse}) and the sum rules
(\ref{eq:m0})--(\ref{eq:m1}), predicts a zero-sound mode that is
significantly too high. This is consistent with the same deficiency of
the Feynman spectrum $\varepsilon(q) = \hbar^2 q^2/2mS(q)$ in \he4.
Drawing on the analogy to \he4 \cite{PinesPhysToday}, this deficiency
can to a large extent be cured by introducing pair fluctuations
$u^{(2)}_{pp'hh'}(t)$ in the excitation operator.

In an approach that maintains the locality of the effective
interactions entering the TDHF equations, we have in
Ref. \onlinecite{2p2h} derived practical working formulas for the
effective interactions in terms of a three-body vertex and a pair
propagator, these are immediately recognized as generalizations of the
bosonic version:
\begin{eqnarray}
V^{(A)}_{ph,p'h'}(\omega)&=& \frac{\delta_{{\bf p}+{\bf h}-{\bf p}'-{\bf h}'}}{N}
\tilde V_{\!_{\rm A}}(q;\omega)\nonumber\\
V^{(B)}_{pp'hh',0}(\omega)&=& \frac{\delta_{{\bf p}+{\bf p}'-{\bf h}-{\bf h}'}}{N}
\tilde V_{\!_{\rm B}}(q;\omega)\,.
\label{eq:VABlocal}
\end{eqnarray}
The components of the (energy--dependent) interactions
$V_{\!_{\rm A, B}}(q;\omega)$ are
\begin{eqnarray}
\tilde V_{\!_{\rm A}}(q; \omega) = \tilde V_{\rm p-h}(q)
&+& [\sigma^{+}_q]^2\, \widetilde W  _{\!_{\rm A}}(q; \omega) +
[\sigma^{-}_q]^2\, \widetilde W^*_{\!_{\rm A}}(q;-\omega)
\\
\tilde V_{\!_{\rm B}}(q; \omega) = \tilde V_{\rm p-h}(q)
&+& \sigma^{+}_q\sigma^{-}_q\, \left[
		\widetilde W_{\!_{\rm A}}  (q; \omega) +
		\widetilde W_{\!_{\rm A}}^*(q;-\omega)
	\right]\;,
\end{eqnarray}
with $\sigma^{\pm}_q \equiv [\SF{q}\pm S(q)]/2S(q)$.  Here, the
$\tilde V_{\rm p-h}(q)$ is the static part of the particle-hole
interaction that is related to the static structure function through
the RPA relationship (\ref{eq:RPAresponse}) and the two sum rules
(\ref{eq:m0}) and (\ref{eq:m1}).
 The energy dependent part of the interaction, $\widetilde W_{\!_{\rm
    A}}(q;\omega)$ describes the splitting and re-combination of phonons,
it consists of a three-phonon vertex $\tilde K_{q,q'q''}$
and a two-phonon propagator $\tilde E^{-1}(q',q''; \omega)$.
The energy dependent correction to the effective interaction is
\begin{equation}
 \widetilde W_{\!_{\rm A}}(q;\omega) =
  \frac{1}{2N}\sum_{{\bf q}',{\bf q}''}\delta_{{\bf q}+{\bf q}'+{\bf q}''}
  |\tilde K_{q,q'q''}|^2 \, \tilde E^{-1}(q',q''; \omega)
\end{equation}
with a three-body vertex
\begin{eqnarray}
\tilde K_{q,q'q''}
&=& \displaystyle \frac{\hbar^2}{2m} \,
	\frac{S(q')S(q'')}{\SF{q}\SF{q'}\SF{q''}}
	\left[
	\qvec\!\cdot\qvec' \, \tilde X_{\rm dd}(q') +
	\qvec\!\cdot\qvec''\, \tilde X_{\rm dd}(q'') - q^2 \tilde{u}_3(q,q',q'')
		\right]\,.
\end{eqnarray}
Here, $\SF{q}$ is the static structure function of non-interacting
fermions.  $\tilde X_{\rm dd}(q) = 1/\SF{q} - 1/S(q)$ can be
identified with the set of ``non-nodal'' diagrams of the
Fermi-hypernetted-chain theory, and ${u}_3(q,q',q'')$ is the
three-body ground state correlation \cite{polish}.

Finally we need the pair propagator:
\begin{eqnarray}
\tilde E^{-1}(q_1,q_2; \omega)
 &=&
 -\!\int\limits_{-\infty}^{\infty}\!\frac{d\hbar\omega'}{2\pi\I}\>
 \kappa(q_1;\omega')\> \kappa(q_2;\omega\!-\!\omega')
\label{eq:Einvqfromkappa}
\\
 \kappa(q;\omega)
 &=&\frac {\kappa_0(q;\omega)}{1+
              \hbar\omega\tilde\Gamma_{\rm dd}(q)\kappa_0(q;\omega)}
\end{eqnarray}
with the ``direct-direct'' correlation function $\tilde\Gamma_{\rm
  dd}(q) = (S(q)-\SF{q})/S_{\rm\scriptscriptstyle F}^2(q)$, and the
partial Lindhard function:
\begin{equation}
\kappa^{\phantom{*}}_0(q;\omega) \equiv \frac{1}{N}\sum_h
\frac{\bar n_{\bf p} n_{\bf h}}{\hbar\omega - e_{ph} + \I\eta}\,.
\end{equation}
Evidently, we need for the execution of the theory only the static
structure function $S(q)$. In this work, we have used input from the
FHNC-EL calculations of Ref. \onlinecite{polish} that were based on
the Aziz-II potential \cite{AzizII}.

\subsection{Bose limit}
\label{ssec:Bose}

For the sake of discussions and to make the connections with previous
work we briefly discuss the Bose limit. In that case, $\SF{q} = 1$,
and we can identify $\tilde\Gamma_{\rm dd}(q) \rightarrow S(q)-1$.
The constituents of the pair propagator
become
\begin{eqnarray}
\kappa_0^{\rm Bose}(q;\omega) &=& \frac{1}{\hbar\omega -t(q)+\I\eta}
\nonumber\\
\kappa^{\rm Bose}(q;\omega) &=& \frac{1}{S(q)}
\frac{1}{\hbar\omega -\varepsilon(q)+\I\eta}\,.
\end{eqnarray}
The pair propagator is then simply
\begin{equation}
\tilde E^{-1}_{\rm Bose}(q_1,q_2; \omega)
= \frac{1}{S(q_1)S(q_2)}\frac{1}{\hbar\omega -\varepsilon(q_1)
-\varepsilon(q_2)+\I\eta}\,.
\label{eq:Ebose}
\end{equation}
The form (\ref{eq:Ebose}) shows more clearly than
Eq. (\ref{eq:Einvqfromkappa}) the physical meaning of the pair
propagator. It also shows the consequence of restricting the
excitation operator to pair fluctuations: The pair propagator contains
in the energy denominator the Feynman spectrum which is known to be
almost a factor of two higher than the physical phonon-roton spectrum.
For bosons, the problem has been cured by introducing multi-particle
fluctuations \cite{QFS09_He4,eomII}, these investigations show that
the energy denominator gets indeed renormalized and closer to the
physical spectrum. We have in the past estimated the effect of
higher-order fluctuations by scaling the Feynman spectrum in the
energy denominator such that the energy of the scaled spectrum agrees
roughly with the experimental roton minimum, {\it i.e.\/} we have used
\begin{equation}
\tilde E^{-1}_\lambda(q_1,q_2; \omega) =
\frac{1}{S(q_1)S(q_2)}\frac{1}{\hbar\omega -\lambda\varepsilon(q_1)
-\lambda\varepsilon(q_2)+\I\eta}
= \frac{1}{\lambda}E^{-1}(q_1,q_2; \omega/\lambda)\,.
\label{eq:Escaled}
\end{equation}
We will use this procedure below to estimate the effect of
multi-particle fluctuations in \he3.

\section{Dynamic structure of $^3$He}
\label{sec:Skw}

We have carried out a comprehensive array of calculations of
$S(q;\omega)$ for \he3 in the experimentally accessible density
regime. Fig. \ref{fig:ezsXray} shows, at saturation density, the
collective mode and the particle-hole continuum corresponding to
single-particle energies of non-interacting particles. Our theoretical
results are compared with X-ray \cite{Albergamo} and neutron
\cite{Fak94} scattering data. The ``zero sound'' mode is, when inside
the particle-hole continuum, identified as the maximum value of
$S(q;\omega)$.

\begin{figure}[ht]
 \includegraphics[width=0.49\textwidth]{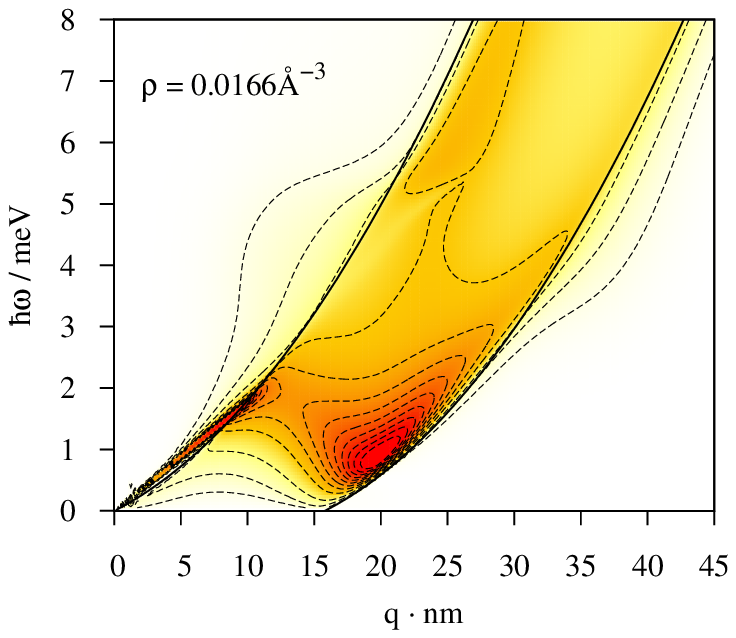}
 \hfill
 \includegraphics[width=0.49\textwidth]{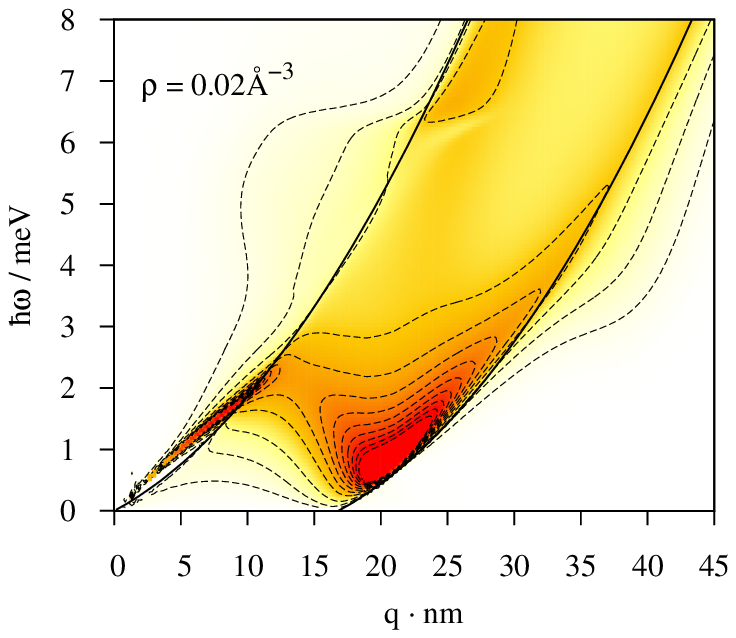}\\
 
 \caption{(Color online) $-\Im m \chi(q;\omega)$ at a density of
   $\rho=0.0166$\AA$^{-3}$ (left pane) and $\rho=0.02$\AA$^{-3}$ (right pane).
The contour lines are at $0.02,\, 0.05,\, 0.1,\,  0.2,\,  0.3,\, 0.4,\,
 0.5,\,  0.6,\, 0.7,\, 0.8,\, 0.9,\, 1.0\, t_F^{-1}$ where $t_F = \hbar^2 k_F^2/2m$ is the Fermi energy of non-interacting \he3 atoms.
 \label{fig:He3_contour}}
\end{figure}

Figs. \ref{fig:He3_contour} give a first glance of our results. The
figures show $S(q;\omega)$ for the densities $\rho= 0.0166$ and $
0.020\,$\AA$^{-3}$, this corresponds to SVP and at a pressure of 10
bar, respectively \cite{GRE86,OuY}.  The figures show the typical
scenario of a collective mode at long wave lengths, and a
particle-hole band. Note, however, that our particle-hole band does
not have sharp boundaries, this effect is due to pair fluctuations
which have the consequence that the interactions $\tilde V_{\!_{\rm
    A,B}}(q; \omega)$ can be complex.

\begin{figure}
 \centerline{\includegraphics[width=0.5\textwidth]{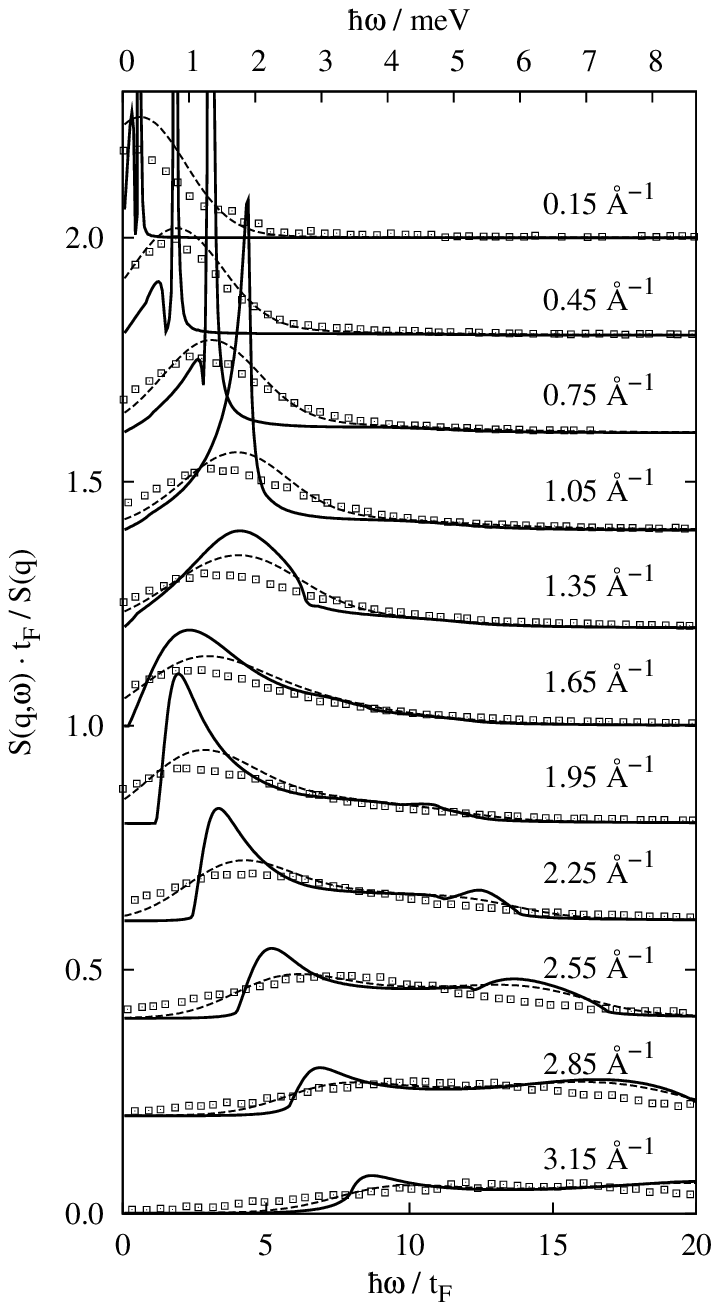}\hfill
             \includegraphics[width=0.5\textwidth]{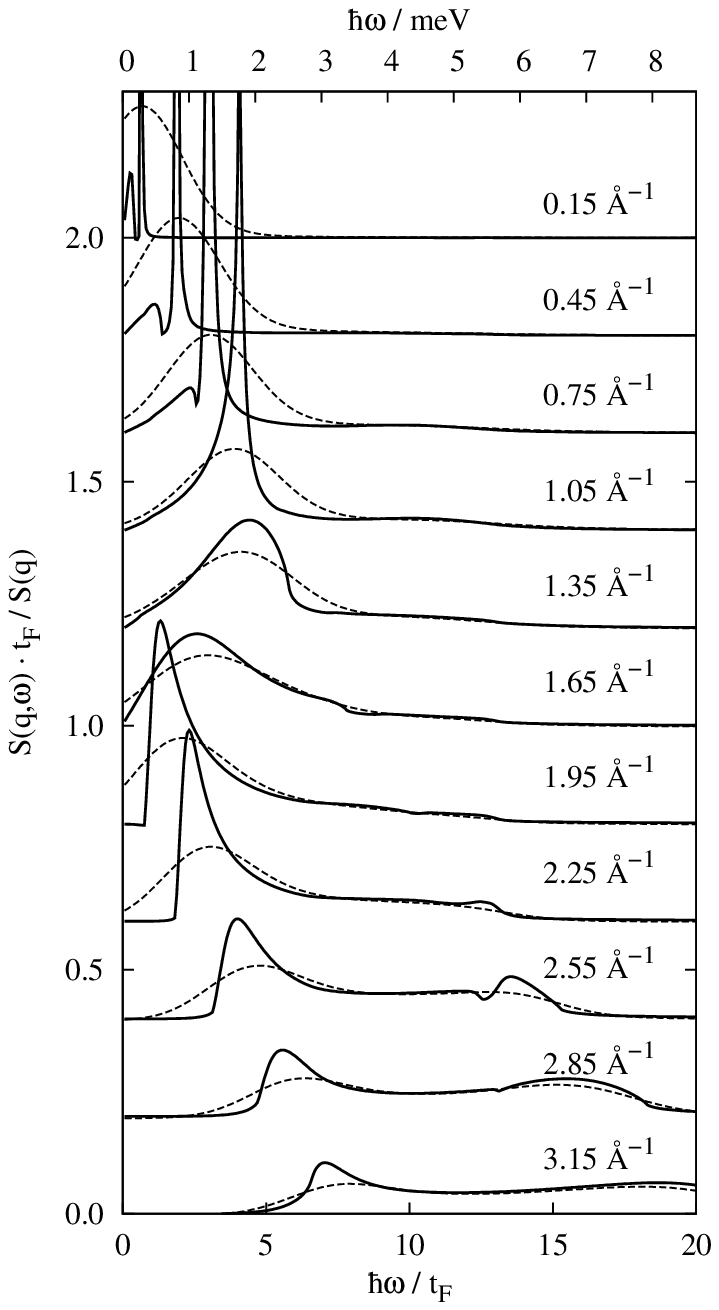}}
\caption{The left pane shows the normalized dynamic structure function
$S(q;\omega)/S(q)$ of \he3 at a density of $\rho=0.0166 $\AA$^{-3}$,
for a sequence of wave numbers. Our theoretical results
(solid line) are compared with the X-ray scattering data
\cite{Albergamo} (squares) at
saturated vapor pressure. Also shown are our theoretical results convoluted
with the experimental resolution (dashed). The right pane
shows the theory at an density of $\rho=0.02$\AA$^{-3}$. All spectra
are normalized to unity. The secondary maximum  observed in Fig.
\ref{fig:He3_contour}
at shorter wave lengths is the due to the possible decay
of excitations into two ``maxon''s and will likely be broadened by
multi-pair fluctuations.
\label{fig:stack}}
\end{figure}

Recent X-ray scattering results \cite{Albergamo} have led to a
discussion on the location of the particle-hole band
\cite{Albergamo_comment,Albergamo_reply}. The authors of
Ref. \onlinecite{Albergamo} state that {\it The obtained results show
  no evidence of such a decay: the zero-sound mode remains well
  defined in the whole explored wave number range.\/} According to
that analysis, the picture shown in Fig. \ref{fig:ezsXray} would not
be correct, rather the particle-hole band should be significantly
lower in energy. In particular the collective mode would lie outside
the particle-hole band (See Fig. 3 of
Ref. \onlinecite{Albergamo}). This is somewhat surprising because such
a scenario would require a single particle spectrum that is
characterized by an average effective mass $m^*\approx 3m$ up to more
than three times the Fermi momentum. This is in contrast to
theoretical work that predicts that the effective mass should have a
peak around the Fermi momentum caused by spin-fluctuations, a
secondary maximum at about twice the Fermi momentum due to density
fluctuations, but then falls off rapidly towards the bare mass
\cite{PethickMass,ZaringhalamMass,Bengt,he3mass}.

Further details are shown in Figs.~\ref{fig:stack}.  To facilitate the
comparison with experiments, we have convoluted our theoretical
spectra with the experimental resolution.  We have scaled our results
by $1/S(q)$ such that the integrated strength is 1 for all momentum
transfers.  We have also scaled the experimental spectra, the scaling
factor agrees, apart from very long wave lengths, with our $1/S(q)$.

Evidently, our results agree, after introducing the broadening, rather
well with the X-ray scattering data. It appears therefore that the
X-ray data can be explained equally well by a scenario where the
collective strength is inside the particle-hole band.  We stress,
however, that such a good agreement can not be obtained in an RPA
model that assumes a form (\ref{eq:RPAresponse}) for the
density-density response function and determines the effective
interaction by the sum rules (\ref{eq:m0}) and
(\ref{eq:m1}). Typically, the strength is distributed over a wider
regime, thus suggesting a broader maximum than observed.

Fig. \ref{fig:stack}b shows the same results for the density of
$\rho=0.020 $\AA$^{-3}$.

\subsection{Multi-pair coupling}

The question arises naturally whether there could be an equivalent to
the ``Pitaevskii plateau'' \cite{Pitaevskii2Roton} in \he3 similar to
one in \he4. In \he4, this plateau is caused by the kinematic
possibility that a ``phonon'' can decay into two ``roton''
excitations. The same process can, of course, also happen for
fermions. However, one expects that the resulting strength is smeared
out because the ``roton'' is {\it per se\/} not a sharp excitation but
can be observed, at its best, by a high density of states in the area
where the roton would be in a Bose system.

To determine if such an effect could be visible we recall our
discussion of the Bose limit in Sec. \ref{ssec:Bose}, in particular
the the consequences of restricting the excitation operator
(\ref{eq:UFermiOft}) to pair fluctuations. The consequence of this
version of the theory is that, in a Bose fluid, the ``plateau''
appears at an energy $\hbar\omega_{\rm pl} = 2 \varepsilon(q_\Delta)$
where $q_\Delta$ is the wave number of the ``Feynman roton'', {\it
  i.e.\/} the minimum of the Feynman dispersion spectrum
$\varepsilon(q)$ \cite{eomI}. We should therefore also expect such a
feature for fermions only at about the energy of the maximum density
of states {\it in the RPA.\/} In both the RPA response function and in
the propagator $\kappa(q;\omega)$ this energy is of the order of 1.4
meV.  The position of the ``roton'' is lowered by pair-fluctuations to
less than 1 meV, {\it cf.\/} Fig. \ref{fig:He3_contour}.

\begin{figure}
 \includegraphics[width=0.49\textwidth]{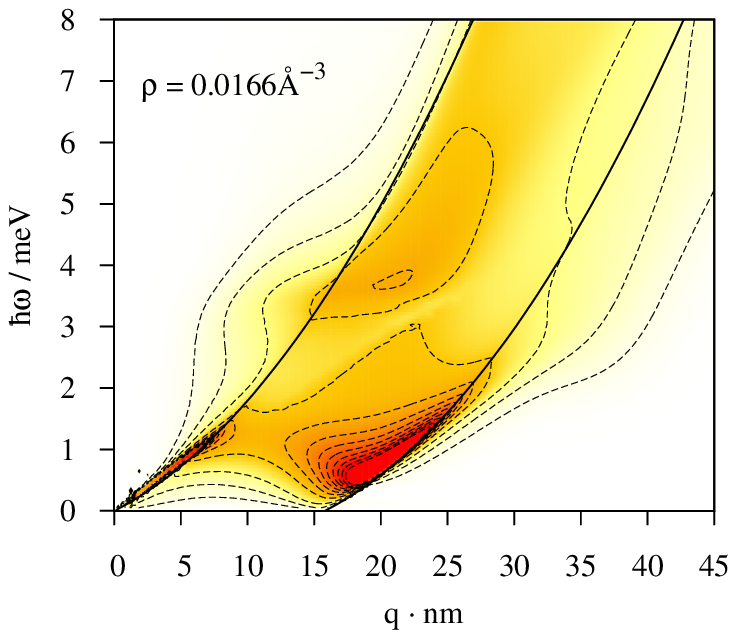}
 \hfill
 \includegraphics[width=0.49\textwidth]{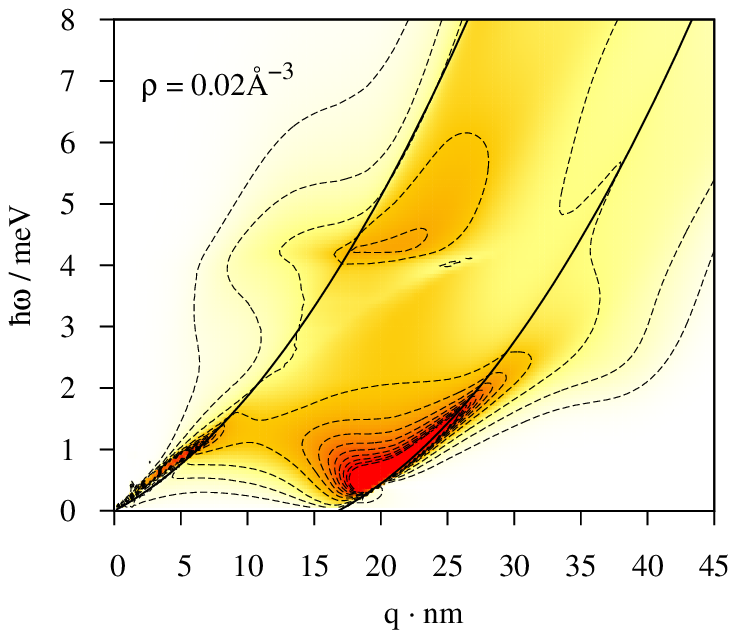}\\
 
 \caption{(Color online) Same as Fig. \ref{fig:He3_contour}
where the pair
propagator is scaled by a factor $\lambda=0.65$ as described in Eq.
(\ref{eq:Escaled})
\label{fig:He3_contour_fudged}}
\end{figure}

In order to estimate the consequence of including multi-particle
fluctuations, we have in previous work \cite{KroQFSBook} scaled the
Feynman spectrum in the pair propagator such that the spectrum in the
energy denominator roughly agrees with the result of the CBF
calculation. Without such a scaling, one would, in the present
calculation, expect a plateau-like structure at an energy of $\hbar
\omega_{\rm pl} \approx 2.8\,$ meV and not at twice the energy of the
actual ``roton''. We adopt here the same philosophy as in the work on
bosons: We scale the two-phonon propagator as formulated in
Eq. (\ref{eq:Escaled}). This modification has little effect for long
wave lengths, but it changes the features of $S(q;\omega)$ for short
wave lengths visibly as seen in Fig. \ref{fig:He3_contour_fudged}. The
most notable effect is the expected broadening of $S(q; \omega)$ at
high momentum transfers.

\begin{figure}
\centerline{\includegraphics[width=8.6cm]{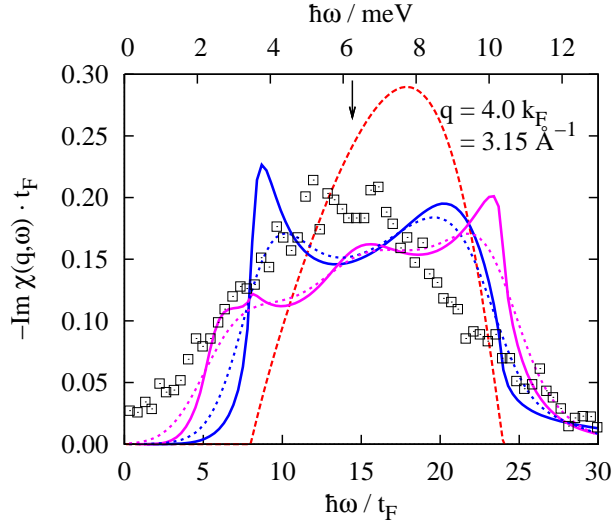}}
\caption{(color online) The figure shows the dynamic structure
function of $^3$He at
saturated vapor pressure at $q=4k_F$. The solid blue line is the
result of the pair excitations theory and dashed red the RPA result.
 The magenta line shows the
result for the pair excitations where the pair propagator has been
modified by a factor $\lambda=0.65$ as described in Eq. (\ref{eq:Escaled}).
The corresponding dashed lines are the theoretical result folded with the
experimental resolution. Inelastic X-ray diffraction data obtained by
Albergamo \textit{et al.}\cite{Albergamo} are also shown (boxes).
\label{fig:plateau}}
\end{figure}

A closeup of the situation is given in Fig. \ref{fig:plateau} where we
show $S(q;\omega)$ at a relatively large momentum transfer
$q=4k_F$. Obviously the more realistic description of two pair
excitations provided by the scaling procedure (\ref{eq:Escaled}) has
the effect of smoothing the spectrum and, in particular, adding
strength to the low-energy part of the spectrum which improves the
agreement between experiment and theory further.

\section{Summary}

We have in this paper presented the application of a new theory of the
dynamic structure function of Fermi systems to \he3. The essential
results have been discussed in the text. Let us reiterate here first
the basic theoretical objectives of our work: The key to a
quantitatively correct description of the dynamics of \he3 -- and,
therefore, other strongly correlated quantum many-body systems -- is
the fact that, for excitations as atomic wave lengths, the {\em
  short-ranged\/} structure of the wave function is time
dependent. The intermediate states can be described only in terms of
the quantum numbers of at least two particles. With that strategy we
have achieved a rather satisfactory agreement between experiment and
theory.

Our results have been obtained without any modification of the
particle-hole spectrum.  Of course, we do {\it not\/} claim that the
precise location of the single-particle spectrum is completely
irrelevant for the energetics of the zero sound mode; but we have
demonstrated that the {\it dominant mechanism\/} in Bose and Fermi
fluids is the same, namely pair-fluctuations.

In independent work \cite{Bengt,he3mass,2dmass}, we have used the
ideas of CBF theory as well as the Aldrich-Pines pseudopotential
theory to calculate the single-particle propagator in \he3.  In both
three and two dimensions, we found the above-mentioned strong
variation of the effective mass with momentum. Near the Fermi surface,
we found good agreement between the theoretical effective mass and the
one obtained experimentally from specific heat measurements.  It is,
however, not clear at all if the simple concept of a single particle
spectrum with a possibly momentum-dependent effective mass is
appropriate away from the Fermi surface. Even simple concepts like the
G(0)W approximation leads to an energy dependent self-energy.  In
order to maintain the sum rules (\ref{eq:m0})--(\ref{eq:m1}), any
modification of the particle-hole spectrum must go along with an
inclusion of exchange effects. At the level of single-pair
fluctuations such a calculation is quite feasible
\cite{NaiThesis,QFS09_exch} but its implementation at the level of the
theory described above still needs to be done.

In terms of the interpretation of experiments we hope that we have
contributed to the clarification of X-ray scattering results. In view
of the large experimental width one is inclined to consider the
theoretical description as satisfactory. At longer wave lengths there
are still some open questions with respect to the onset of Landau
damping \cite{Fak94} which we hope to clarify in the future by the
inclusion of the above-mentioned self-energy corrections.

\begin{acknowledgments}
  A part of this work was done while one of us (EK) visited the
  Physics Department at the University at Buffalo, SUNY. Discussions
  with H.\ M.\ B\"ohm, C. E. Campbell H. Godfrin, R.\ Holler, and
  R. E. Zillich are gratefully acknowledged. This work was supported,
  in part, by the Austrian Science Fund FWF under project P21264.
\end{acknowledgments}

\newpage
\bibliography{papers}
\bibliographystyle{apsrev}
\end{document}